\documentclass[11pt]{article}
\usepackage{times}
\usepackage{geometry}
\usepackage[utf8]{inputenc}
\usepackage{enumitem,amssymb}
\usepackage{graphicx}
\usepackage{fancyhdr}
\usepackage{aas_macros}
\usepackage{mdframed} 
\usepackage{hyperref}

\usepackage[authoryear]{natbib}
\setcitestyle{authoryear,open={(},close={)}}

\mdfdefinestyle{theoremstyle}{
innertopmargin=\topskip,}
\mdtheorem[style=theoremstyle]{lrptextbox}{}

\title{Probing Diverse Phenomena through Data-Intensive Astronomy}

\author{Mubdi Rahman, Dunlap Institute, University of Toronto; \\ Dustin Lang, Perimeter Institute;\\ Ren\'ee Hložek, Dept. of Astronomy \& Astrophysics, Dunlap Institute, University of Toronto;\\ Jo Bovy, Dept. of Astronomy \& Astrophysics, Dunlap Institute, University of Toronto; \\
Laurence Perreault-Levasseur, Department of Physics, Université de Montréal and MILA, Montréal;}
\date{}
\begin{document}
% ****BEGIN MAIN WHITE PAPER SECTION****

\maketitle

% Instructions: Please insert your white paper text here.
%A white paper should be a self-contained description of a future opportunity for Canadian astronomy. A white paper will be most effective and useful if it concisely summarises and recommends an option that the LRP2020 panel should be considering for prioritisation.
%
% White papers are not required to contain a specific set of sections or headings. Depending on the content, the following topics may be appropriate to include:
%connection or relevance to Canada
%timeline 
%cost 
%description of risk
%governance / membership structure 
%justification for private submission of supplementary information

% The full document can have a maximum length of 10 pages including text, figures, tables, responses to selection criteria, references and appendices, and a size of 30 MB.

\begin{abstract}

The era of data-intensive astronomy is being ushered in with the increasing size and complexity of observational data across wavelength and time domains, the development of algorithms to extract information from this complexity, and the computational power to apply these algorithms to the growing repositories of data. Data-intensive approaches are pushing the boundaries of nearly all fields of astronomy, from exoplanet science to cosmology, and they are becoming a critical modality for how we understand the universe. The success of these approaches range from the discovery of rare or unexpected phenomena, to characterizing processes that are now accessible with precision astrophysics and a deep statistical understanding of the datasets, to developing algorithms that maximize the science that can be extracted from any set of observations. 

In this white paper, we propose a number of initiatives to maximize Canada's ability to compete in this data-intensive era. We propose joining international collaborations and leveraging Canadian facilities for legacy data potential. We propose continuing to build a more agile computing infrastructure that's responsive to the needs of tackling larger and more complex data, as well as enabling quick prototyping and scaling of algorithms. We recognize that developing the fundamental skills of the field will be critical for Canadian astronomers, and discuss avenues through with the appropriate computational and statistical training could occur. Finally, we note that the transition to data-intensive techniques is not limited to astronomy, and we should coordinate with other disciplines to develop and make use of best practises in methods, infrastructure, and education. 
  
\end{abstract}

\section{Astronomy in a Data-intensive Era}
\label{sect:dataintensiveera}

Astronomy is entering the era of data-intensive science. This transition is being driven by the greater size and complexity of the data being generated by new facilities, as well as the development of algorithms with the ability to extract and make sense of this complexity, along with the computational power to make use of these algorithms. The data-intensive era does not simply refer to the size of typical datasets produced, but rather the complexity that may be extracted from such datasets; while current and future survey programs have reasonable initial challenges that come from the need to record and store such data, the growth of computational resources tends to catch up \citep{schlegel2012lsst}. Rather, we are referring to are the opportunities and challenges that arise with increased algorithmic and data complexity. 

Data-intensive approaches have permeated nearly all fields of astronomy, enabling the discovery of rare or unexpected phenomena, exploring processes that are only accessible through precision astrophysics, measuring parameters that require large populations with well-understood statistics, and developing algorithms to optimally extract information within the growing complexity of astronomical data, often independent of their immediate applicability. These collection of approaches have produced some of the most impactful discoveries over the past decade. We highlight some of the successful approaches to data-intensive astronomy and results that have arisen below. 

\subsubsection*{Discovering Rare or Unexpected Phenomena}

A natural avenue of discovery in the era of larger, complex datasets is identifying rare or unexpected phenomena. This could be phenomena that have been predicted but previous datasets were insufficient to contain a single example of them, or phenomena that were altogether unpredicted but are serendipitously captured in a sufficiently large dataset. Among the most notable example of the latter is the discovery of fast radio bursts \citep{2007Sci...318..777L}, neither predicted nor expected, which was discovered through archival Parks radio telescope data. The discovery kicked off the development of new transient radio searches \citep[e.g.][]{2018ApJ...863...48C}. In contrast, an example of a phenomena that had been predicted, but has only been found through data-intensive techniques is that of tidal disruption events. These events with distinct time-varying signatures have been primarily identified through random, non-targeted observations in the X-ray \citep[e.g.][]{2015ApJ...811...43L} or the optical \citep[e.g.][]{2013MNRAS.430L..45A}. It is important to note that in both of these examples, the precise control of the cadence and statistics of the observations was not initially critical, since the existence of the signal alone is sufficient to generate new science, independent of initial estimates of rate or prevalence. 

In the case of more intentional survey datasets, the discoveries of rare and unexpected phenomena have been plentiful; for instance, the use of the Wide-field Infrared Survey Explorer to identify populations of circumstellar disks, crucial to understanding the early stages of the formation of planets \citep{2016ApJ...830...84K}. The Dark Energy Survey has been used to identify populations of strongly-lensed quasars, which can act as laboratories for a variety of phenomena related to gravitation, cosmology, and galaxy formation \citep{10.1093/mnras/stz2200}. Closer to home, that survey has been used to discover a population of RR Lyrae in the Galactic Halo \citep{2019AJ....158...16S}, which can be used as tracers for substructure. The Apache Point Observatory Galactic Evolution Experiment provides another example, where an extreme metal-poor field giant has been found with conflicting kinematic and chemical properties, identifying mysteries in how the disk stellar population was built \citep{2016ApJ...833..132F}.

In addition to the science that is being developed from offline surveys or archival observations, immediate or near-real time phenomena identification is becoming crucial, especially in the field of time-domain astrophysics. As new, deeper, high-cadence facilities come online, such as the Large Synoptic Survey Telescope \citep{2009arXiv0912.0201L}, real-time classification becomes critical as the number of transient phenomena identified will greatly exceed the ability for follow-up. For instance, there are efforts on how to best type supernovae photometrically to maximize follow-up potential \citep[e.g.][]{2011MNRAS.414.1987N}. This challenge has even pushed the community to think creatively and look beyond astronomy; for instance the PLAsTiCC challenge as an example, which used a crowd-sourced competition approach to develop transient classification schemes \citep{2019PASP..131i4501K, 2019arXiv190704690B}. Solving the problem of real-time transient classification is crucial to the success of time-domain astronomy with the upcoming generation of facilities. 

\subsubsection*{Enabling Precision Astrophysics}

The increasing size and statistical power of new datasets is enabling science that had previously been impossible. This has led to discoveries arising from untangling high-dimensional datasets, such as the work of understanding the Milky Way's chemical cartography with the APOGEE survey \citep{2015ApJ...808..132H}, to using the very same spectra as backlights to infer the physical properties of the mysterious diffuse interstellar bands \citep{2015ApJ...798...35Z}. These results come from well-defined surveys with characterizable selection functions, helping discern small-scale systematic trends from statistical noise. 

Coordinating surveys across multiple wavelength regimes increases the information available to beyond the sum of their component parts. For example, this has been leveraged to infer the physical properties of circumgalactic gas around galaxies \citep{2014ApJ...795...31L}, essentially impossible to do without information across multiple wavelengths. Adding the time-domain as another dimension further increases the complexity of the dataset, and along with it, increases its ability to describe the underlying physical processes it traces. For example, the use of high-cadence data to constrain supernova shock breakouts \citep{2016ApJ...832..155F}. 

Beyond just complexity, the larger size of these surveys have further enabled better understanding and control of the systematics of the observations both the catalogue level, and the pixel level. Understanding the systematics inherent in the catalog construction has enabled such results as finding candidates for the most luminous OB associations in the Milky Way from the Two-Micron All Sky Survey \citep{2013ApJ...766..135R}, and identifying statistically ``weird'' galaxies in the Sloan Digital Sky Survey \citep{2017MNRAS.465.4530B}. Characterizing the systematics at the imaging level increases the likelihood of even detecting sources that were not visible in the initial pass, for instance, reprocessing the Legacy survey to identify Ultra-faint galaxies \citep{2019ApJS..240....1Z}. It is through the combination of increased complexity of datasets, as well as a more thorough understanding of the systematics that go into their construction, that precision astrophysics is possible. 

\subsubsection*{Discovery through Parameter Inference}

The era of data-intensive astronomy has enabled novel methods to measure or infer the values of otherwise inaccessible parameters. These can include cross-correlation methods between datasets of different wavelengths \citep[e.g. inferring the cosmic star formation rate from Planck;][]{2015MNRAS.446.2696S}, or of sets with different known properties \citep[e.g. inferring the clustering redshift of objects in a photometric catalog;][]{2016MNRAS.460..163R}. In the field of exoplanets, there have been breakthroughs in inferring the planet population statistics from catalogues with poor constraints on their underlying statistics \citep{2014ApJ...795...64F}. Another important process in astronomy is labelling and classifying sources, and these process have seen breakthroughs in the data-intensive era. For example, the classification of stellar parameters has seen increased accuracy and efficiency through new approaches \citep{2015ApJ...808...16N}. 

Beyond larger, more complex data sets, parameter inference techniques are being applied to existing datasets, such as the case of Bayesian technique being used to infer the stellar birth function of stellar clusters \citep{2015ApJ...808...45G}. The techniques that are being developed for the large, high-complexity datasets of the future are being used to extract the maximal amount of information out of the legacy datasets of today. 

\subsubsection*{Algorithm Development}

A less visible but growing area of research is the development of algorithms and techniques to conduct data-intensive astronomy. This research area is significantly interdisciplinary, with links between all fields of astronomy to computer science, statistics, and other data-intensive sciences. With the complexity of datasets outpacing the growth in computing power we can use to explore them, algorithm development has become as critical to data-intensive astronomy as instrument development has been to classical observational astronomy. 

An example of such an area is the development of techniques to extract the maximal amount of information from image pixel data; The UnWISE algorithms and catalogues are examples of the processes that can be used to increase the value of space-based data for an entire survey \citep{unwise}. Similarly, the development of graph and Bayesian techniques have improved the photometry from confused images from the Herschel Space Observatory \citep{2015ApJ...798...91S}. 

The algorithms developed in service of astronomy have had impact outside of the field, especially in the area of Bayesian methods; algorithms and their implementations such as emcee \citep{2013PASP..125..306F} and MultiNest \citep{2009MNRAS.398.1601F} have seen extensive use both inside and outside astronomy. These efforts provide an example of the types of interdisciplinary collaboration that could be led by the astronomical community, with direct benefits felt beyond our field.  

\subsection*{The Challenges and Opportunities of Data-Intensive Astronomy}

It is important to note that many of these results would not have been possible without access to the pre-existing datasets that were observed; they either required a sufficiently large dataset to identify rare events or phenomena, or have come serendipitously through an exploration of the data. Attempting to get the necessary observing resources through traditional PI-driven processes would be improbable, as time allocation committees tend to down-weight proposals where the perceived probability of success is low, or the resources required to conduct such a study are seen as excessive in comparison to their potential scientific payoff. To continue the growth of data-intensive astronomy, priority must be given to systematically designed multipurpose datasets. 

Astronomy is not the only field going through this transition right now. Other fields in the natural sciences and beyond, including chemistry, biology, health sciences, genomics, and political sciences are facing this same embarrassment of riches on similar timescales. While exact details of computational and data problems will not be the same from field-to-field, questions of infrastructure, training, and community coordination may be shared or analogous. This provides an opportunity for the astronomical community to learn from the example of other fields. 

\section{Positioning Canada in Data-Intensive Astronomy}

\subsection{Access to Datasets and Surveys}

Canadian astronomers already have access to immense numbers of complex datasets, either through specific access agreements, or due to their open access policies. In the case of open access datasets and surveys, individual members of the Canadian community have been involved with projects using {\em GAIA}, {\em SDSS} public data, {\em WMAP}, {\em PanSTARRS}, and many more. Through access agreements or partnerships, some Canadian astronomers have access to propietary datasets such as those from {\em Planck} and {\em SDSS}, but are limited to individual researchers or institutions. Canadian will have some degree of participation in the future of large survey programs, such as {\em LSST} and {\em Euclid}, but the nature of access and support that Canadian astronomers receive from these projects will depend strongly on what the community's buy-in will be, both in terms of financial contribution and personnel dedicated to their development and operations. 

For proprietary surveys, such as {\em SDSS-V}, buy-in is a necessary part of operations and access, and ensures a seat at the table for what scientific goals are prioritized. Beyond science access, early buy-in is critical to developing the necessary expertise to allow the community to maximize its investment. Even for surveys with a mandate for open access, we strongly recommend that Canada use its personnel and financial resources to enter collaborations while survey design and operations are still being planned. 

It is important to note that open access to data does not imply equal access, especially when considering the activation energy required to make use of a dataset at an in-depth level. Even for fully public projects, it is important that the community develops local expertise or has access to trained expertise. For ``full-service'' collaborations, there are models for what the delivery of this support looks like (e.g., CASjobs and the underlying support at SDSS, the North American ALMA Science Centres). Developing or buying into these support structures for larger public surveys should be a part of the Canadian long-term strategy.   

\subsection{Leveraging National Facilities}

Canada has national access to a variety of ground-based and space-based facilities, across multiple wavelength regimes. Canadian facilities have traditionally focused on PI-driven proposals with limited time spent on nationally coordinated legacy programs. However, using more of the existing access for larger, coordinated legacy programs with shared scientific goals would have many benefits. This includes the ability to professionalize/systematize data reduction procedures and pipelines, minimizing any individual risk pool of failed observations by sharing across the program, and providing value for the Canadian-investment in the facility beyond its funding timescale. In fact, such an operational mode would even enhance the efficiency of PI-led programs through better developed shared tools, techniques, pipelines, and expertise developed from the legacy programs. Historically, Canada has had successes with legacy data products, for instance the Canadian Galactic Plane Survey has had 30 citations in 2017 alone, a decade-and-a-half after its data release in 2003. 

In the landscape of Canadian astronomy, a number of facilities are primed for legacy programs; both the {\em Canada-France-Hawaii Telescope} and the {\em Gemini Observatory} have significant national access, and instruments that would provide unique datasets if used in survey mode. Further, the right legacy programs may interest other international partners, opening the possibility to share time allocations across multiple partners. On the space-based side, Canadian access to {\em Astrosat}, in particular, would be invaluable in a legacy program being one of the few operational ultraviolet facilities at present. Enabling legacy programs on these facilities would require policy changes at the Canadian Time Allocation Committee and the Canadian Space Agency level, but the potential benefit to researchers across Canada, and the impact of this Canadian access on long-term science output would be significant.   

Even in the case of PI-driven use of Canadian observatories, these datasets can be useful in a legacy context if the appropriate systems exist to capture it when no longer proprietary, assess the data quality, reduce it with a consistent pipeline, and provide it with an accessible interface. There are existing examples of such systems, for instance the Chandra Source Catalog \citep{2010ApJS..189...37E}, and the Hubble Source Catalog \citep{2016AJ....151..134W}. In both cases, the catalogs provide a uniform data product derived from heterogeneous observations across multiple instruments from PI-led targeted projects. Data curation in systems like this should be the default for any data taken with a national facility. 

In addition to national facilities,  the instrumentation community in Canada provides another avenue to create and lead unique legacy programs by leveraging Canadian-built instruments. Through designing legacy programs based around such instruments, Canada can provide niche, impactful data products while boosting the scientific output of the designed instrument. The example of {\em CHIME} shows how, with an appropriate niche carved out, even modest cost facilities can provide an impactful legacy dataset.

\section{What infrastructure do we need?}

\subsection{Computational Infrastructure}

A critical necessity for conducting data-intensive astronomy is access to sufficient computational resources. In Canada, computation resources are acquired in one of two ways; purchased for individual researchers/research groups through research grants, or accessed through shared facilities via Compute Canada.  Individual research grants will rarely be sufficient to purchase the necessary computing, and this mode of delivering resources is be wasteful since computer utilization by an individual group will generally be below full capacity. Consequently, access to shared facilities is the primary mode that Canadian researchers can obtain sufficient computational resources. 

A concern with the current structure, primarily provided through Compute Canada, is that both the hardware available and the process through which time is attained are insufficiently agile for data-intensive astronomy. The typical national computing cluster is optimized for traditional high-performance computing applications, where the emphasis is minimizing latency in interconnects to optimize highly-parallel algorithms. These facilities are excellent for many traditional computational astrophysics programs, such as hydrodynamic simulations, but add punishing overheads to even the simplest of data-intensive astronomy problems. Additionally, resources are often allocated based on the presumption of predictable, steady computational usage and modes (i.e., specific codes running on the same cluster continuously for some expected amount of time until convergence). This acts as a barrier towards rapid test-and-refinement method that is a necessary part of phenomenologically-driven data-intensive research.

Traditional HPC facilities tend to focus on processor or GPU power rather than rapid-access storage, which is needed for algorithmic development. The system of per-user or per-group storage allocation make it difficult for any given user to host and/or manage any reasonably large dataset, and induces inefficiency where multiple users may have copies of the same datasets. This makes it difficult to keep your data close to your computing, which is a bottleneck in many data-intensive approaches. Developing structures that allow large, common datasets to be shared amongst multiple members of the community would allow for a more efficient use of resources, as well as lowering the overhead for researchers to use the facility. 

Finally, the support provided by national computing facilities tends to be focused on traditional computing challenges, such as parallelization and load balancing, which makes it difficult to test and adopt new developments in data-intensive techniques. For instance, many data-intensive techniques are being developed in or make use of databases \citep[e.g.][]{microsoft_2017}, which require technical support to run and maintain. Reorienting the available support towards shared computational resources for the astronomical community would allow researchers to be more agile in their discovery science and their algorithm development. 

There are initiatives that are beginning to bridge these gaps between support provided by the national facilities and the needs of the data-intensive astronomy community. Notably, the Canadian Initiative for Radio Astronomy Data Analysis (CIRADA) is funded to develop science-ready data products for the Canadian radio community. The Canadian Advanced Network for Astronomical Research (CANFAR) is developing user-facing systems to deliver data-intensive astronomy and interface with Compute Canada. Additionally, there are proposals to build data and software centres for the Canadian LSST initiatives. 

\subsubsection*{The Rise of Cloud Computing}

In addition to dedicated research computing facilities, the development of virtualization and cloud computing technologies is enabling new ways to interact with data; these systems allow users to rapidly prototype, refine, and scale algorithms and analyses. There are systems set up to leverage these technologies for users, including within academic computing (i.e., the SciServer project), and through public commercial cloud computing providers (i.e., Amazon AWS, Microsoft Azure, Google Cloud). Increasing the access that Canadian researchers have to cloud and virtualization technologies would be a boon for data-intensive astronomy, either through having them hosted at national facilities, or coordinating pools of credits for commercial cloud providers. 

\subsection{Support Infrastructure}

While data access to all the growing numbers of surveys is an absolute minimum, it is insufficient as the data rights alone do not provide expertise, or any influence in the design or science requirements of the survey. Hosting personnel who work on the functional aspects of the surveys is critical to the Canadian community, in particular to build the sufficient institutional expertise to leverage the data that we suddenly have access to. These individuals may act as instructors for scientists who have potential ideas of how to approach a particular dataset, as well as evangelists for the potential uses of the dataset across multiple fields. In particular, funding personnel to work on functional aspects of future surveys enables the Canadian community to ``buy-in'' to a larger number of opportunities using existing or more readily-available funding opportunities, such as CFI grants. CFI funding has already been succesfully leveraged to develop computational and support infrastructure in the form of CIRADA for radio astronomy. As the community develops this support infrastructure, national coordination and knowledge-sharing becomes critical to the efficiency and efficacy of these systems. Consequently, we recommend having a forum through which this coordination can occur.  

\section{How do we develop the Canadian astronomical community for this era?}

The paradigm of data-intensive science is novel for many astronomers, and internalized conceptual barriers are often just as detrimental as infrastructure or access barriers. As a community, we need to support building familiarity with large-dataset approaches, which requires institutional and personnel support in the same way the use of a new instrument or observing facility would. This may take the form of support personnel who have in depth knowledge of the data sets, their availability, and are in functional positions to help non-specialist researchers leverage these products. Additionally, it will require algorithmic specialists who interface with the computer science and statistics communities to develop suites of tools to maximize the science output from the available datasets.   

\subsection{Education and Training}

A crucial aspect of developing a robust data-intensive astronomy community in Canada is the training of students, and rethinking the basics of astronomy undergraduate and graduate education. While the traditional astronomy education places physics and mathematics training as the fundamentals, this new era of astronomy demands that formal computational and statistical training becomes a requirement. The current landscape of statistical training for Canadian astronomers and recommendations is discussed in the LRP2020 whitepaper on Astrostatistics (E017). There are growing efforts to deliver the required training on various fronts; for instance, the LSST Data Science Fellows program delivers an intense course of training for graduate students on the technical skills necessary to approach the science questions the survey seeks to address. Additionally, international conferences and summer schools on astroinformatics and astrostatistics have acted as professional development venues for active researchers. 

A question arises to which vehicle is best to deliver such training. Astronomy departments have not necessarily developed the experience to provide this kind of training, nor are Computer Science and Statistics departments sufficiently specific or motivated to develop the skills that are required for data-intensive astronomy. However, astronomy is not the only field making this transition now, especially with the rise of computational biology, chemistry, and other fields. So, a possible solution may be a hybrid solution to the necessary computational and statistical training or courses. 

A useful example may be in the field of Public Health, where statistical fluency is a necessary fundamental within the field. Rather than depending on courses from statistics departments, especially at the graduate level, these fundamentals are often developed in purpose-driven biostatistics courses. Importantly, the syllabi of these courses are built such that the tools developed in them can be used in concurrent courses within the curricula. This model may work well in astronomy.

\subsection{Interdisciplinary Interaction}

The Canadian astronomy community is not in this transition to data-intensive science alone; how to use of large, complex data products and increasingly diverse algorithms is a question being posed as a part of the Canadian National AI strategy \citep{canadian_ai_strategy}. The astronomical community should be active as a part of these discussions as the relevant development programs, expertise, and support networks that will be set up for other fields may inform similar structures we seek to develop in our own. This can be one of the ways in which our relatively small research community can get access to much larger facilities and resources. 

% ****BEGIN CRITERIA SECTION (4 page limit) ****

%Instructions:
%Assessment and prioritisation of facilities and programs in LRP2020 will be based on a predefined set of criteria. 
%Authors are requested to explicitly address these criteria in the set of text boxes below. Some criteria may not be applicable to all white papers. 
% IMPORTANT: 
% There is no specific length limit on individual boxes. 
% However, the full set of 8 boxes should comprise no more than 4 pages and these pages **do** count toward the 10-page limit of the full document.

\begin{lrptextbox}[How does the proposed initiative result in fundamental or transformational advances in our understanding of the Universe?]

There is a great need for infrastructure development to support large astronomical data sets and to train the next generation in data science and analysis to maximize the science potential of next-generation instruments. These datasets touch on multiple science interests and by supporting their construction, we will be able to make transformational advances in our understanding of fundamental questions about our universe.

\end{lrptextbox}

\begin{lrptextbox}[What are the main scientific risks and how will they be mitigated?]
The risks are that the large data volumes produced by upcoming telescope facilities go unused or underused. Given the data volume and rate, much of this data will need to be pre-processed, and so a lack of planning will reduce the impact of such large datasets.
    
\end{lrptextbox}

\begin{lrptextbox}[Is there the expectation of and capacity for Canadian scientific, technical or strategic leadership?] 
Canada has a growing capacity for leadership in this area, partly due to the scientific leadership of previous large surveys and the significant impact of current large-data projects like CHIME. This proposal would leverage this existing expertise.

\end{lrptextbox}

\begin{lrptextbox}[Is there support from, involvement from, and coordination within the relevant Canadian community and more broadly?] 
There is broad support form the Canadian community for data-intensive astronomy. This support must be galvanised to fund, support and maintain the data infrastructure needed to drive innovation in large data science. 

\end{lrptextbox}

\begin{lrptextbox}[Will this program position Canadian astronomy for future opportunities and returns in 2020-2030 or beyond 2030?] 
Ensuring access to and analysis of large data projects will bring great rewards for Canadian astronomy in the next decade, which will surely be the decade of big data astronomy.

\end{lrptextbox}

\begin{lrptextbox}[In what ways is the cost-benefit ratio, including existing investments and future operating costs, favourable?] 
Building in sufficient costs for infrastructure support and data management has not been the strong suit of the astronomical community in Canada and beyond. Firm software and computation support for large telescope facilities is now being prioritized as the only way to ensure these projects and the data they produce have longevity.

\end{lrptextbox}

\begin{lrptextbox}[What are the main programmatic risks
%Instructions: Programmatic risks include but are not limited to schedule, feasibility, budget, technical readiness level, computational or software requirements, dependence on other partners, and governance plan.
and how will they be mitigated?] 

The programmatic risks are to continue in an `old' modality where the data are simply stored in a server and analyzed locally by astronomers. Such a failure of planning will leave Canadian astronomy behind in the fully online analysis mode that is expected of large future surveys like LSST, CHIME and the SKA.

\end{lrptextbox}

\begin{lrptextbox}[Does the proposed initiative offer specific tangible benefits to Canadians, including but not limited to interdisciplinary research, industry opportunities, HQP training,
%HQP=Highly qualified personnel, defined as individuals with university degrees at the bachelors' level and above.
EDI,
%EDI = equity, diversity and inclusion 
outreach or education?] 
%insert your text here 
Starting a national conversation about coordinated funding scheme for analysis (similar to e.g. the ADAP funding cycle purely to support analysis of archival data) will also provide tangible training for the next generation in terms of skills development and analysis. The needed computational and statistical analysis have the potential to be interdisciplinary and are naturally transferable to data-intensive careers outside of astronomy.
\end{lrptextbox}

\setlength{\bibsep}{0.0pt}
\bibliography{example} 

\end{document}